\documentclass[12pt,a4paper]{article}

\usepackage{epsfig,graphicx}
\usepackage{amsmath,amsfonts,amssymb}
\newcommand{\overbar}[1]{\mkern 1.5mu\overline{\mkern-1.5mu#1\mkern-1.5mu}\mkern 1.5mu}

\newcommand{\unit}{\leavevmode\hbox{\small1\kern-3.6pt\normalsize1}}

\def\lsim{\raise0.3ex\hbox{$\;<$\kern-0.75em\raise-1.1ex\hbox{$\sim\;$}}}
\def\gsim{\raise0.3ex\hbox{$\;>$\kern-0.75em\raise-1.1ex\hbox{$\sim\;$}}}

\def    \beq            {\begin{equation}}
\def    \eeq            {\end{equation}}
\def    \bea           {\begin{eqnarray}}
\def    \eea           {\end{eqnarray}}

\def \mn{\mu\nu{\rm SSM}}

\def\g2{{\rm GeV}^2}
\begin{document}

\thispagestyle{empty}
\begin{flushright}
FTUAM 17/1\\
IFT-UAM/CSIC-17-001\\
\vspace*{5mm}
\today
\end{flushright}

\begin{center}
{\Large \textbf{ 
On a reinterpretation of the Higgs field in supersymmetry 
and a proposal for new quarks
}
}

\vspace{0.5cm}
\hspace*{-1mm}
D.E.~L\'opez-Fogliani$^{a,b}$  and C.~Mu\~noz$^{c,d}$\\[0.2cm]
{$^{a}$\textit{IFIBA, UBA \& CONICET, Departamento de F\'{\i}sica, 
FCEyN,\\ Universidad de Buenos Aires, 1428 Buenos Aires, Argentina}\\}
{$^{b}$\textit{Pontificia Universidad Cat\'olica Argentina, Buenos Aires, Argentina}\\}
{$^{c}$\textit{Departamento de F\'{\i}sica Te\'{o}rica, Universidad Aut\'{o}noma de Madrid,\\
Campus de Cantoblanco, E-28049 Madrid, Spain}\\}
{$^{d}$\textit{Instituto de F\'{\i}sica Te\'{o}rica UAM-CSIC,\\ Campus de Cantoblanco, E-28049 Madrid, Spain}}\\[0pt]

\begin{abstract}
In the framework of supersymmetry, when $R$-parity is violated the Higgs doublet superfield $H_d$ can be interpreted as another doublet of leptons, since all of them have the same 
quantum numbers. Thus Higgs scalars are sleptons and Higgsinos are leptons.
We argue that this interpretation can be extended to the second Higgs doublet superfield $H_u$, when right-handed neutrinos are assumed to exist.
As a consequence, we advocate that this is the minimal construction where the two Higgs doublets can be interpreted in a natural way as a fourth family of lepton superfields, and that this is more satisfactory than the usual situation in supersymmetry where the Higgses are `disconnected' from the rest of the matter and do not have a three-fold replication.
On the other hand, in analogy with the first three families where for each lepton representation there is a quark counterpart, we propose a possible extension of this minimal model including  
a vector-like quark doublet representation as part of the fourth family. We also discuss the phenomenology of the associated new quarks.
\end{abstract}
\end{center}

\vspace*{15mm}\hspace*{3mm}
\hspace*{28mm} new quarks.
\newpage


%
%

\section{Introduction}

The Higgs particle in the framework of the standard model is intriguing, being the only elementary scalar in the spectrum, and introducing the hierarchy problem in the theory. Besides, whereas for the rest of the matter there is a three-fold replication, this does not seem to be the case of the Higgs since only one scalar/family has been observed.
In the framework of supersymmetry,
the presence of the Higgs is more natural: 
scalar particles exist by construction, the hierarchy problem can be solved, and the models predict that the Higgs mass must be 
$\lsim$ 140 GeV if perturbativity of the relevant couplings up to high-energy scales is imposed. In a sense, the latter has been confirmed by the detection of a scalar particle with a mass of about 125 GeV.
However, in supersymmetry the existence of at least two Higgs doublets, $H_d$ and $H_u$, is necessary, as in the case of the minimal supersymmetric standard model (MSSM) \cite{martin}, and as a consequence new neutral and charged scalar particles should be detected in the future to confirm the theory.
Similar to the standard model, no theoretical explanation is given for the existence of only one family of Higgs doublets.


In this work we want to contribute a new vision of the Higgs(es) in the framework of supersymmetry.
We will argue that the well known fact that the Higgs doublet superfield $H_d$ has the same gauge quantum numbers as the doublets of leptons $L_i$, where $i=1,2,3$ is the family index,
is a clue that the Higgses can be reinterpreted as a fourth family of lepton superfields. Thus Higgs scalars are sleptons and Higgsinos are leptons. 
This can be done only when $R$-parity ($R_p$) is violated, since 
the standard model particles and their superpartners have 
opposite $R_p$ quantum numbers.
Early attempts in this direction can be found in Refs.~\cite{kaku,chun}. In particular, in the first paper it was pointed out that in theories with TeV scale quantum gravity, the scalar $H_d$ can be a fourth family slepton. Since $H_u$ is not present in that construction, with its role in the Lagrangian played by $H_d$ through non-renormalizable couplings, $H_d$ is proposed to be part of a complete standard model family in order to cancel anomalies. In the second paper, in the context of low-energy supersymmetry the scalar
$H_u$ was also included as a slepton as part of another complete family with opposite quantum numbers to the fourth family.
Thus, four chiral families with standard model quantum numbers and one chiral family with opposite quantum numbers are present in that construction.

However, with the matter content of the MSSM,
which is sufficient to cancel anomalies, this interpretation 
of $H_d$ as another lepton superfield in the case of $R_p$ violation cannot be extended to $H_u$ in a natural way,
as we will show in Section~2.
Fortunately, as we will discuss in Section~3, when right-handed neutrino superfields are allowed in the spectrum, not only the violation of $R_p$ turns out to be natural solving the $\mu$ problem and reproducing easily current neutrino data, but also the interpretation of $H_u$ as part of the fourth family of lepton superfields is straightforward. Finally, we will argue in Section~4 that, as a consequence, a vector-like quark doublet representation might also be part of the new fourth family, and we will briefly discuss its phenomenology. 
Our conclusions are left for Section~5.

\section{Supersymmetry without right-handed neutrinos}

Unlike the standard model where only one Higgs doublet scalar 
(together with its complex conjugated representation)
is sufficient to generate Yukawas couplings for quarks and charged leptons at the renormalizable level, in supersymmetry we need 
a vector-like Higgs doublet representation, with their superfields usually denoted as:
\bea
H_d = \begin{pmatrix} H^{0}_d\\
H^{-}_d
\end{pmatrix},\:  
{H_u} = \begin{pmatrix} {H^{+}_u}\\
{H^{0}_u}
\end{pmatrix}.
\label{fieldcontent-Hsm}
\eea
In addition,
the matter sector of the supersymmetric standard model, in the absence of right-handed neutrinos, 
contains also the following three families of superfields:
\bea
\:\:\:\:\:\:\:\:\:\:\:\:\:\:\:\:\:\:\:\:\:\:\:\:\:\:\:\:\:\:\:\:\:\:\:\:\:\:\:\:
L_{i}=\begin{pmatrix} \nu_{i}\\
e_{i}
\end{pmatrix},\:\:\:\:\:\:\:\:\:
\begin{array}{c}
e_{i}^{c}\\
-
\end{array},\:\:\:\:\:\:\:\:\:\:\:
Q_{i}=\begin{pmatrix} u_{i}\\
d_{i}
\end{pmatrix},\:
\begin{array}{c}
d_{i}^{c}\\
u_{i}^{c}
\end{array},
\label{fieldcontent-Lsm}
\eea
where 
we have defined 
$u_i$, $d_i$, $\nu_i$, $e_i$, 
and $u^c_i$, $d^c_i$, $e^c_i$, 
as the 
left-chiral superfields whose fermionic components are the left-handed fields
of the corresponding quarks, leptons, and antiquarks, antileptons, respectively. 

With this matter content,
the most general gauge-invariant renormalizable superpotential is given by:
\bea\label{superpotentialtotal}
W
&=& 
\mu
\, {H}_u \,  H_d  
+
Y^e_{ij} \, H_d\, L_i \, e_j^c
+
Y^d_{ij} \, H_d\, Q_i \, d_j^c 
-
Y^u_{ij} \, {H}_u\,  Q_i \, u_j^c
\nonumber\\
&
+
&
\mu_i
{H}_u \, L_i  
+
\lambda_{ijk} \, L_i  \, L_j \,  e_k^c
+
\lambda'_{ijk} \, L_i \, Q_j \, d_k^c
+
\lambda''_{ijk} \, u_i^c\, d_j^c\,  d_k^c
\,,
\eea
where the summation convention is implied on repeated indexes, and
our convention for the contraction of two $SU(2)$ doublets is e.g.
${H}_u \,  H_d\equiv  \epsilon_{ab} H^a_u \,  H^b_d$,
with $\epsilon_{ab}$ the totally antisymmetric tensor $\epsilon_{12}=1$.

In the absence of the terms in the second line,
the terms in the first line of Eq.~(\ref{superpotentialtotal}) 
constitute the superpotential of the MSSM,
where baryon ($B$) and lepton ($L$) numbers are conserved.
This superpotential arises from imposing the $Z_2$ discrete symmetry $R$-parity \cite{barbier}, 
$R_p=(-1)^{2S}(-1)^{(3B+L)}$, which acts on the components of the superfields. Here 
$S$ is the spin, and one obtains $R_p=+1$ for ordinary particles and $-1$ for their superpartners.
Because of the different $R_p$ quantum numbers, there can be no mixing between particles and superpartners.



If we allow the terms in the second line of Eq.~(\ref{superpotentialtotal}) to be present, they
violate $R_p$ explicitly \cite{barbier}.
The first term $\mu_i{H}_uL_i$ which also violates lepton number, together with the superpotential of the MSSM, constitute 
the bilinear $R$-parity violation model (BRpV). 
This term 
contributes to the neutral scalar potential generating 
VEVs not only for the Higgses as in the MSSM, but also for the left sneutrinos,
$\langle \widetilde\nu_{iL} \rangle\neq 0$.
The other three terms are the conventional trilinear lepton- and baryon-number-violating couplings.
The presence of the couplings $\mu_i, \lambda_{ijk},\lambda'_{ijk}$,
violating lepton number
could have easily been argued, once the $\mu$-term and the Yukawa couplings 
for $d$-type quarks and charged leptons are introduced in the first line of
the superpotential (\ref{superpotentialtotal}), by noting  
that the superfields
$H_d$ and $L_i$ have the same gauge quantum numbers.
Actually, the latter fact might lead us to interpret the
Higgs superfield $H_d$ as a fourth family of lepton superfields $L_4$, in addition to the three families $L_i$ of Eq.~(\ref{fieldcontent-Lsm}):
\bea
L_{4}=\begin{pmatrix} \nu_{4}\\
e_{4}
\end{pmatrix} 
= \begin{pmatrix} H^{0}_d\\
H^{-}_d
\end{pmatrix} = H_d\,.
\label{ele4}
\eea
Notice that this is not possible in the case of the MSSM because the components of the superfields $H_d$ and $L_i$ have opposite quantum numbers under $R_p$.
Unfortunately, we cannot interpret naturally the other Higgs superfield 
${H}_u$ in a similar way, given that it has no leptonic counterpart, in particular its neutral component.  
We will see in the next section that this counterpart is present when we enter right-handed neutrinos in our supersymmetric framework.

On the other hand, it is well known that the simultaneous presence of the couplings 
$\lambda'_{ijk}$ and $\lambda''_{ijk}$, violating lepton and baryon number respectively,
 can be dangerous since they would produce fast proton decay. The usual assumption in the literature of the MSSM of invoking 
$R_p$ 
to avoid the problem is clearly too stringent, since then the other 
couplings $\lambda_{ijk}$, and $\mu_i$ in the superpotential
(\ref{superpotentialtotal}), which are harmless for proton decay, would also be forbidden.
A less drastic solution, taking into account that the choice of 
$R_p$ 
is {\it ad hoc},
is to use other $Z_N$ discrete symmetries to forbid only
$\lambda''_{ijk}$. This is the case e.g. of $Z_3$ Baryon-parity~\cite{Ibanez:1991hv} which also prohibits dimension-5 proton decay operators, unlike $R_p$.
In addition, this strategy seems reasonable if one expects all discrete symmetries to arise from the breaking of gauge symmetries of the underlying unified theory~\cite{krauss}, 
because
Baryon-parity and 
$R_p$ are the only two generalized parities which are `discrete gauge' anomaly free~\cite{Ibanez:1991hv}.
Discrete gauge symmetries are also not violated~\cite{krauss} by potentially dangerous
quantum gravity effects~\cite{gilbert}. 

Given the relevance of string theory as a possible underlying unified theory, a robust argument in favour of the above mechanism is that, in string compactifications such as e.g. orbifolds,
the matter superfields have
several extra $U(1)$ charges broken spontaneously at high energy by the Fayet-Iliopoulos D-term, and as a consequence residual $Z_N$ symmetries 
are left in the low-energy theory.
As pointed out in Ref.~\cite{analisisparam},
the same result can be obtained by the complementary mechanism 
that stringy selection rules can naturally forbid the $\lambda''_{ijk}$ couplings discussed above, since matter superfields are located in general in different sectors of the compact space.
As a whole, some gauge invariant operators violating $R_p$ can be forbidden, but others are allowed~\cite{orbifolds}.

Let us finally remark that although the BRpV has the interesting property of generating through the bilinear terms $\mu_i$ that mix the left-handed neutrinos 
$\nu_{iL}$ and the neutral Higgsino $\tilde H^0_u$, one neutrino mass at tree level (and the other two masses at one loop), the $\mu$ problem \cite{mupb} is in fact augmented with the three new supersymmetric mass terms which must be  
$\mu_i\lsim 10^{-4}$ GeV, in order to reproduce the correct values of neutrino masses.
This extra problem can be avoided imposing a $Z_3$ symmetry in the superpotential, which implies that only trilinear terms are allowed.
Actually, this is what one would expect from a high-energy theory where
the low-energy modes should be massless and the massive modes of the order of the high-energy scale.
As pointed out in Ref.~\cite{propuvSSM}, this is what happens in 
string constructions,
where the massive modes have huge masses of the order of the string scale and the massless ones have only trilinear terms at the renormalizable level. Thus one ends up with an accidental $Z_3$ symmetry in the low-energy theory.

To summarize the discussion, instead of the superpotential of 
Eq.~(\ref{superpotentialtotal}), a more natural superpotential (in the sense of free of  problems) 
with the minimal matter content of Eqs.~(\ref{fieldcontent-Hsm}) and (\ref{fieldcontent-Lsm}) seems to be
\bea\label{superpotentialnatural}
W
&=& 
Y^e_{ij} \, H_d\, L_i \, e_j^c
+
Y^d_{ij} \, H_d\, Q_i \, d_j^c 
-
Y^u_{ij} \, {H_u}\, Q_i \, u_j^c
\nonumber\\
&
+
&
\lambda_{ijk} \, L_i  \, L_j \, e_k^c
+
\lambda'_{ijk} \, L_i \, Q_j \, d_k^c
\,.
\eea
However, 
this implies that not only the bilinear terms $\mu_i$ are forbidden, 
but also the crucial $\mu$ term generating Higgsino masses.
In the next section we will discuss a solution through the presence of right-handed neutrinos which will also allow us to interpret the two Higgs doublet superfields as a fourth family of leptons.

\section{
Right-handed neutrinos and reinterpretation of the Higgs superfields
}

Right-handed neutrinos are likely to exist in order to generate neutrino masses.
Thus we will add 
these superfields 
to the minimal matter content of Eq.~(\ref{fieldcontent-Lsm}), allowing us also to write it in a more symmetric way:
\bea
L_{i}=\begin{pmatrix} \nu_{i}\\
e_{i}
\end{pmatrix},\:
\begin{array}{c}
e_{i}^{c}\\
\nu_{i}^{c}
\end{array},\:\:
Q_{i}=\begin{pmatrix} u_{i}\\
d_{i}
\end{pmatrix},\:
\begin{array}{c}
d_{i}^{c}\\
u_{i}^{c}
\end{array}\,.
\label{fieldcontent-L}
\eea
This spectrum together with the Higgs superfields in Eq.~(\ref{fieldcontent-Hsm}) give rise to the following gauge invariant superpotential proposed in Refs.~\cite{propuvSSM,analisisparam}:
\bea\label{superpotential1}
W &=&  
Y^e_{ij} \, H_d\, L_i \, e_j^c
\ +
Y^d_{ij} \, H_d\, Q_i \, d_j^c 
\ -
Y^u_{ij} \, {H_u}\, Q_i \, u_j^c
\ -
Y^\nu_{ij} \, {H_u}\, L_i \, \nu^c_j
\nonumber\\
&+&
\lambda_{ijk} \, L_i  \, L_j \, e_k^c
+
\lambda'_{ijk} \, L_i \, Q_j \, d_k^c
+
\frac{1}{3}
\kappa_{ijk}\, \nu^c_i \, \nu^c_j \, \nu^c_k
+\lambda_{i} \, {H_u} \, H_d \, \nu^c_i
\,.
\eea
This superpotential expands the one in Eq.~(\ref{superpotentialnatural}) with the right-handed neutrinos, and is built using the same arguments of Section~2 in order to forbid the bilinear terms $\mu$ and $\mu_i$, and the couplings violating baryon number $\lambda''_{ijk}$ of superpotential (\ref{superpotentialtotal}). 
The first line corresponds to Yukawa couplings which conserve $R_p$, whereas the couplings in the second line violate $R_p$ explicitly.
In the absence of $\lambda_{ijk}$ and $\lambda'_{ijk}$, $R_p$ is restored in the limit
$Y^\nu \rightarrow 0$.

The three terms with couplings
$Y^\nu$, $\lambda_i$ and $\kappa$
are characteristics of the
`$\mu$ from $\nu$' supersymmetric standard model ($\mn$) 
\cite{propuvSSM,reviewsmunu3}, and are harmless for proton decay. 
They contribute to the neutral scalar potential generating 
VEVs not only for the Higgses and the left sneutrinos as in the BRpV, but also for the right sneutrinos. 
As a consequence of the electroweak symmetry breaking, 
a $\mu$ term  of the order of the electroweak scale is generated dynamically with $\mu=\lambda_i \langle \widetilde\nu_{iR} \rangle^*$. Let us also remark that the term with
$Y^\nu$ contains
the Dirac Yukawa couplings for neutrinos, 
and besides generates effectively bilinear couplings, 
$\mu_i=Y^\nu_{ij}\langle \widetilde\nu_{jR} \rangle^*$, as those discussed in Section~2 for the BRpV.
The $\kappa$ term produces Majorana masses for the right-handed neutrinos, 
$M_{ij}=\sqrt 2\kappa_{ijk}\langle \widetilde\nu_{kR}\rangle^*$,
instrumental in the generation of correct neutrino masses and mixing through a generalized electroweak-scale seesaw 
\cite{propuvSSM,analisisparam,Ghosh:2008yh,Bartl:2009an,javier,Ghosh:2010zi,LopezFogliani:2010bf}. 

Because of the VEVs acquired by the neutral scalars and the violation of $R_p$, all fields in the spectrum with the same color, electric charge and spin mix together contributing to the rich phenomenology of the $\mn$.
For example, 
the neutral (scalar and pseudoscalar) Higgses mix with the left and right sneutrinos,
the charged Higgses with the charged sleptons, the
neutralinos of the MSSM with the left- and right-handed neutrinos,
and the charginos with the charged leptons.
Besides, in the $\mn$ the scale of the breaking is set up by the soft terms, which is in the
ballpark of a TeV. This nice features give rise to realistic signatures of this model at colliders~\cite{Ghosh:2008yh,Bartl:2009an,ghosh,fidalgo,probing,hunting,probing2}, well verifiable at the LHC or at upcoming accelerator experiments. 
For example, prompt and/or displaced multi-leptons/taus/jets/photons final states.

Concerning cosmology in the $\mn$,
as a consequence of $R_p$ violation the lightest supersymmetric particle (LSP) is no longer a valid candidate for cold dark matter. Nevertheless, embedding the model in the context of supergravity, one can accommodate the 
gravitino~\cite{choi,reviewsmunu3} as an eligible decaying dark matter candidate with a lifetime greater than the age of the Universe. Its detection is also possible through the observation of a gamma-ray line in the Fermi satellite~\cite{choi,clues,fermi,fermi2}. In Ref.~\cite{chung}, the generation of the baryon asymmetry of the universe was analysed in the $\mu\nu$SSM, with the interesting result that electroweak baryogenesis can be realised. 


Similarly to the discussion for  
$\lambda_{ijk}$ and $\lambda'_{ijk}$ in Section~2,  
the presence of the
$\lambda_i$ term which violates lepton number in the superpotential (\ref{superpotential1}), 
could have been deduced from the presence of the couplings $Y^{\nu}$
because of the same quantum numbers for the superfields
$H_d$ and $L_i$.
Actually, the simultaneous presence of both terms in order to solve the $\mu$ problem and generate correct neutrino masses implies that all the charges of $H_d$ and $L_i$ must be the same even if extra $U(1)$'s are present.
Thus the argument of the extra $U(1)$ charges used in Section~2 to forbid the couplings
 $\lambda_{ijk}''$ is unlikely that can be used in typical string models to forbid the couplings $\lambda_{ijk}$ and $\lambda'_{ijk}$. This makes more robust the superpotential (\ref{superpotential1}).
Besides, as discussed in Ref.~\cite{analisisparam}, even if $\lambda_{ijk}$ and $\lambda'_{ijk}$ are set to zero, they are generated by loop corrections (although with very small values) due to the presence in the superpotential of couplings like $Y^d, Y^\nu, \lambda_i$.

In Section~2, the fact that the superfields $H_d$ and $L_i$ have the same gauge quantum numbers
led us to discuss in Eq.~(\ref{ele4}) the possibility of interpreting 
$H_d$ as a fourth family of lepton superfields $L_4$.
However, we were not able to interpret naturally the Higgs superfield
${H_u}$ in a similar way, given that it has no leptonic counterpart in the spectrum of Eq.~(\ref{fieldcontent-Lsm}).  
On the contrary, for the spectrum of Eq.~(\ref{fieldcontent-L}) it is possible to interpret ${H_u}$
as another lepton superfield $L^{^c}_4$:
\bea
L^{^c}_4
=
\begin{pmatrix} e_{4}^{c}\\
\nu^c_{4}
\end{pmatrix} 
= 
\begin{pmatrix}
{H^{+}_u}
\\
{H^{0}_u}
\end{pmatrix} 
= 
{H_u}\,.
\label{ele4barra2}
\eea
Thus, at the level of weak eigenstates the superfield $H_d/L_4$ contains the fourth-family left sneutrino and the $H_u/L_4^c$ the fourth-family right sneutrino, as shown in Eqs.~(\ref{ele4}) and~(\ref{ele4barra2}).
In the limit were the others sneutrinos are decoupled in our model, the Higgs discovered at the LHC is described by a mixture of $H_u$ and $H_d$ as in the case of the MSSM.  In addition, also as in the latter case, for reasonable values of 
$\tan\beta$ the standard model-like Higgs is mainly $H_u$. Therefore, in this supersymmetric framework the first scalar particle discovered at the LHC is mainly a right sneutrino belonging to a fourth-family vector-like lepton doublet representation.

To complete the argument, we must take into account what was mentioned above, that once the electroweak symmetry is broken the first three families of sneutrinos turn out to be mixed with the fourth one. Nevertheless, the left sneutrinos of the first three families are decoupled in all cases, since the mixing occurs through terms proportional to neutrino Yukawas or left sneutrino VEVs which are very small~\cite{analisisparam}.
Concerning the right sneutrinos of the first three families, they are singlets of $SU(2)$ and can mix in general with the doublets $H_u$ and $H_d$, similarly to the case of the Next-to-MSSM (NMSSM)~\cite{nmssm} where one extra singlet is present.
As a consequence, the decoupling limit is not necessarily a good approximation. For our model, where three singlets are present, discussions about viable regions of the parameter space and the expected signals at colliders were carried out in Refs.~\cite{fidalgo} and~\cite{probing2}. In those works, where not only LHC constraints but also LEP and Tevatron ones were applied to the parameter space, viable regions were obtained.

Summarizing, Eqs.~(\ref{ele4}) and~(\ref{ele4barra2}) constitute our reinterpretation of  
Eq.~(\ref{fieldcontent-Hsm}), and therefore we can write the whole spectrum in the following way:
\bea
\:\:\:\:\:\:\:\:\:\:\:\:\:\:\:\:\:\:\:\:\:\:\:\:\:\:\:\:\:\:\:\:\:\:\:
L_{i}=\begin{pmatrix} \nu_{i}\\
e_{i}
\end{pmatrix},\:\:\:\:\:\:\:\:\:\:\:
\begin{array}{c}
e_{i}^{c}\\
\nu_{i}^{c}
\end{array},\:\:\:\:
Q_{i}=\begin{pmatrix} u_{i}\\
d_{i}
\end{pmatrix},
\begin{array}{c}
d_{i}^{c}\\
u_{i}^{c}
\end{array},
\nonumber
\label{fieldcontent-SMtotal2}
\eea
\bea
L_{4}=\begin{pmatrix} 
\nu_{4}\\
e_{4}
\end{pmatrix} 
,\:  
L^{^c}_4
=
\begin{pmatrix} e_{4}^{c}\\
\nu^c_{4}
\end{pmatrix}.
\label{fieldcontent-total2}
\eea
With this notation, Eq.~(\ref{superpotential1}) can be written in a more compact way as:
\begin{equation}\label{superpotential-originalcompact}
W
= 
\ Y^e_{IJk} \, L_I \, L_J \, e_k^c +
Y^d_{Ijk} \, L_I \, Q_j \, d_k^c 
-
Y^u_{4jk} \, L^{^c}_4
\, Q_j  \, u_k^c -
Y^\nu_{4Jk} \, L^{^c}_4
\, L_J \, \nu^c_k
+
\frac{1}{3}
\kappa_{ijk}  \, \nu^c_i  \, \nu^c_j  \, \nu^c_k
\,,
\end{equation}
where $I=i,4$ 
and $J=j,4$ are 
the new family indexes, with $i,j,k=1,2,3$, and the notation for the Yukawa couplings is self-explanatory.

\section{Proposal for new quarks
}

We have identified in the previous section the minimal model where the two Higgs superfields can be interpreted in a natural way as a fourth family of leptons.
One might think that this is just an academic discussion, in the sense that 
superpotential (\ref{superpotential-originalcompact}) is equivalent from the
operational viewpoint to superpotential (\ref{superpotential1}).
Nevertheless, in this framework where in principle vector-like matter can be added to the fourth family consistently with the experiments, we find it natural to make the following proposal.
In analogy with the first three families in Eq.~(\ref{fieldcontent-total2}), where each lepton representation has its
quark counterpart, we add to the spectrum of the fourth family a vector-like quark doublet representation
as counterpart of the vector-like lepton/Higgs doublet representation, implying in superfield notation:
\bea
L_{i}=\begin{pmatrix} \nu_{i}\\
e_{i}
\end{pmatrix},\:\:\:\:\:\:\:\:\:\:\:\:\:\:\:\:\:
\begin{array}{c}
e_{i}^{c}\\
\nu_{i}^{c}
\end{array},\:\:\:\:\:\:\:\:
Q_{i}=\begin{pmatrix} u_{i}\\
d_{i}
\end{pmatrix},\:\:\:\:\:\:\:\:\:\:\:\:\:\:
\begin{array}{c}
d_{i}^{c}\\
u_{i}^{c}
\end{array},
\nonumber
\label{fieldcontent-SMtotal}
\eea
\bea
L_{4}=\begin{pmatrix} \nu_{4}\\
e_{4}
\end{pmatrix} 
,\: 
 L^{^c}_4 =
\begin{pmatrix} e_{4}^{c}\\
\nu^c_{4}
\end{pmatrix},\: 
Q_{4} = \begin{pmatrix} u_{4}\\
d_{4}
\end{pmatrix},\:\:
Q^{^c}_4=
\begin{pmatrix} d_{4}^{c}\\
u_{4}^{c}
\end{pmatrix}\,,
\label{fieldcontent-total}
\eea
where $Q_4$ has hypercharge $\frac{1}{6}$ as for the first three families, whereas 
$Q^{^c}_4$
has by construction hypercharge $-\frac{1}{6}$ allowing the cancellation of 
anomalies\footnote{Other extensions of the $\mn$ 
were discussed in Ref.~\cite{Fidalgo:2011tm} in the context of an extra $U(1)$ gauge symmetry.}.
It is worth noticing here that the presence of extra vector-like matter is a common situation in string constructions\footnote{For a standard-like model containing only the Higgs doublets as vector-like representations, see however Ref.~\cite{marchesano}. Remarkably, in that model the presence of three families of right-handed neutrinos is mandatory to achieve anomaly cancellation.}
(see e.g.~\cite{orbifolds,casas,orbifolds2,shiu}).

The spectrum of Eq.~(\ref{fieldcontent-total}) implies that 
the following terms associated to the presence of the new quarks
must be added to the superpotential in Eq.~(\ref{superpotential1}):
\begin{equation}\label{extratermss2}
W
= 
\lambda'_{i4k} \, L_i\, Q_4 \, d_k^c +
Y^d_{4k} \, H_d\, Q_4 \, d_k^c
-
Y^u_{4k} \, {H_u}\, Q_4 \, u_k^c +
Y_{j4k}^Q \, Q_j  \, 
Q^{^c}_4
\,  \nu_k^c
+
Y_{44k}^Q \, Q_4  \, 
Q^{^c}_4
\,  \nu_k^c
\,,
\end{equation}
where the first one corresponds to trilinear lepton-number-violating couplings, the second and third contribute to the Yukawa couplings with the Higgses, and the last two terms contribute to the quark masses once the right sneutrinos acquire VEVs. 

Working in low-energy supersymmetry,
these terms 
will induce the corresponding trilinear soft-supersymmetry breaking terms in the Lagrangian.
Together with the soft masses for the 
squark doublets $\widetilde Q_4$ and $\widetilde Q^{^c}_4$
they constitute the new terms in the soft Lagrangian. Notice that none of them contributes to the minimization of the tree-level neutral scalar potential.

Using now the `new' notation for the Higgs superfields, 
Eq.~(\ref{extratermss2}) 
can be written as
\begin{equation}\label{extratermss3}
W
= 
 Y^d_{I4k} \, L_I\, Q_4 \, d_k^c
-
Y^u_{44k} \, 
L^{^c}_4
\, Q_4 \, u_k^c +
Y_{J4k}^Q \, Q_J \, 
Q^{^c}_4
\,  \nu_k^c
\,.
\end{equation}
This equation together with Eq.~(\ref{superpotential-originalcompact}) allow us to write the whole superpotential in the compact notation:
\begin{align}\label{superpotential3}
W
= &
\ Y^e_{IJk} \, L_I \, L_J \, e_k^c +
Y^d_{IJk} \, L_I \, Q_J \, d_k^c 
-
Y^u_{4Jk} \, 
L^{^c}_4
\, Q_J \, u_k^c 
-
Y^\nu_{4Jk} \, 
L^{^c}_4
\, L_J \, \nu^c_k
+
Y_{J4k}^Q  \, Q_J \,  
Q^{^c}_4
\,  \nu_k^c
\nonumber\\
& 
+
\frac{1}{3}
\kappa_{ijk}  \, \nu^c_i  \, \nu^c_j  \, \nu^c_k
\,.
\end{align}

New phenomenology is expected from the presence of the new quarks (and squarks) of the fourth family.
Here we will discuss the specially interesting case 
of the quarks, given their mixing with the standard model ones and therefore the modification of the usual couplings to the $W$, $Z$ and Higgs boson.
For example, although in this construction the Higgs mass is already enhanced at tree level due to the $\lambda_i$ couplings~\cite{analisisparam},
the presence of the new quarks could help to enhance it further through one-loop effects~\cite{staub}.
Besides, the presence of flavour changing neutral currents (FCNCs) leads to a wide range of final states that can be analysed.
Notice that large enough masses for the new quarks
to be beyond the present experimental bounds, but still accessible at the LHC, can be generated by the last term in Eq.~(\ref{extratermss2}) with a Yukawa coupling 
$Y^Q_{44k}\sim 1$ and typical VEVs of the right sneutrinos 
$\langle \widetilde\nu_{kR} \rangle \sim$ TeV as discussed in Section~3.


In the basis of 2-components spinors $(u_L^*)^T=\left(u_{IL}^*\right)$, $(u_R)^T=\left(u_{JR}\right)$, one obtains the following 
up-quark mass terms in the Lagrangian:
\begin{equation}
{\cal L}_{\text{mass}}= -(u_{L}^*)^T
{m}_{u} u_R + \mathrm{h.c.}\, ,
\label{matrixupquarks}
\end{equation}
where, using a compact block notation, 
\begin{align}
m_u=
\left(\begin{array}{cc}
{Y^u_{ij}}^*   
\langle H_u^0\rangle^*
& 
{Y^{Q}_{i4k}}^*\langle \widetilde v_{kR}\rangle
\\
{Y^{u}_{4j}}^*
\langle H_u^0\rangle^*
& 
{Y^Q_{44k}}^*\langle \widetilde v_{kR}\rangle
\end{array}\right)\,.
\label{submatrix}
\end{align}
We can simplify further this matrix redefining the left-handed fields in such a way that the new entries
$(m_u)_{i4}$ are vanishing and 
$(m_u)_{ij}={Y'^u_{ij}}^*\langle H_u^0\rangle^*$, with  $Y'^u$ the redefined Yukawa coupling.
After these replacements, the $4\times 4$ mass matrix is diagonalized by two unitary matrices $U_L^u$ and $U_R^u$:
\begin{equation}
{U_L^u}^{^{\dagger}} m_{u} {U_R^u} = m_{u}^{\text{dia}}
\, ,
\label{diagmatrixdownquarks}
\end{equation}
with
\begin{eqnarray}
u_R = {U_R^u} U_R\,,\,\,\,
u_L = {U_L^u} U_L\,.
\label{physdownquarks}
\end{eqnarray}
Here, the 4 entries of the matrices $U_L$, $U_R$ are the 2-component up-quark mass eigenstate fields.
After a phase redefinition of the $d_{4R}$ field to recover the conventions for the non-supersymmetric standard model extensions with vector-like 
quarks~\cite{aguilar2}, 
the same formulas apply to the down-quark sector with the replacements
$Y^u \rightarrow Y^d$ and
$\langle H_u^0\rangle \rightarrow \langle H_d^0\rangle$
in Eq.~(\ref{submatrix}).

Taking the above mixing matrices into account, in the basis of 4-components spinors with the projectors $P_{L,R}=\frac{1}{2}(1\mp\gamma_5)$, charged currents are modified in the following way:
\begin{eqnarray}
{\cal L}_{W} &=&  
 -\frac{g}{\sqrt 2}\left(
\overbar U\gamma^{\mu} V_L
P_L D
+
\overbar U\gamma^{\mu} V_R
P_R D
\right) W^+_\mu
+ \mathrm{h.c.}
\,,
\label{chargedcurrents}
\end{eqnarray}
where 
\begin{eqnarray}
V_L
= {U_L^u}^{^{\dagger}} U_L^d
\,,\,\,\,\,\,
V_R
= {U_R^u}^{^{\dagger}} \delta_R^{\text{dia}} U_R^d
\,,
\label{ckmL}
\end{eqnarray}
and the matrix $\delta_R^{\text{dia}}=\text{dia} (0,0,0,1)$.
Here, the measured CKM matrix corresponds to the (non-unitary) $3\times 3$ block
$(V_L
)_{ij}$, and another (non-unitary) CKM matrix $V_R$ must be defined for the right-handed quarks because of the new doublet
$Q_{4R}^T=(u_{4R}, d_{4R})$.

Tree-level FCNC also occur due to the mixing in the right-handed sector induced by the new doublet.
In particular, 
the neutral current interactions of quarks in the Lagrangian are:
\begin{equation}
{\cal L}_{Z} =  
 -\frac{g}{
2\text{cos}\,\theta_W
}
\left(
\overbar U\gamma^{\mu} P_L U
+
\overbar U\gamma^{\mu}
X^u
P_R U
-
\overbar D\gamma^{\mu}  P_L D
-
\overbar D\gamma^{\mu}  
X^d
P_R D
-2\sin^2\theta_W J^\mu_{em}
\right)Z_\mu
\,,
\label{neutralcurrents}
\end{equation}
where the matrices
\begin{equation}
X^u=V_R V^{^{\dagger}}_R  = {U_R^u}^{^{\dagger}}\delta_R^{\text{dia}} U_R^u\,,\,\,\,\,\,
X^d=V^{^{\dagger}}_R  V_R  = {U_R^d}^{^{\dagger}}\delta_R^{\text{dia}} U_R^d\,,
\end{equation}
are hermitian and non-diagonal, mediating FCNCs.

Finally, the modified couplings between the neutral components of the Higgs doublets and quarks
\begin{eqnarray}
{\cal L}_{H_u} &=&  
 -\frac{1}{\langle H_u^0\rangle^*}
\overbar U
\left(m_{u}^{\text{dia}}
-(m_u)_{44}{U_L^u}^{^{\dagger}}  \delta_R^{\text{dia}} 
U_R^u
\right) P_R U\ {H_u^0}^*
+ \mathrm{h.c.}
\,,
\nonumber\\
{\cal L}_{H_d} &=&  
 -\frac{1}{\langle H_d^0\rangle^*}
\overbar D
\left(m_{d}^{\text{dia}}
-(m_d)_{44}{U_L^d}^{^{\dagger}}  \delta_R^{\text{dia}} 
U_R^d
\right) P_R D\ {H_d^0}^*
+ \mathrm{h.c.}
\,,
\label{higgscurrent}
\end{eqnarray}
may change the production and decay of the standard model-like Higgs.

Numerous analyses of the phenomenology of vector-like quark singlet, doublet and triplet representations have been carried out in the literature in extensions of the standard model~\cite{aguilar,okada,aguilar2},
studying limits on mixing between ordinary quarks and heavy partners, the allowed range of splitting between the heavy states,
and the production at the LHC. 
In particular, the non-observation of FCNCs put stringent constraints on mixing, and only one light quark can have significant mixing with the vector-like quark.
Since the vector-like quarks are usually expected to mix predominantly 
with the third generation~\cite{aguilar3}, one can obtain upper limits on the corresponding mixing angles from new contributions to the oblique parameters
$S$ and $T$, and $Z\rightarrow b\bar b$ observables~\cite{aguilar2}. These limits for the case of a vector-like doublet $(T,B)$, where $T$ is a new up-type of quark with charge +$2/3$ and
$B$ is a new down-type of quark with charge $-1/3$, can be applied to our model and are given for the right-handed fields by $\sin\theta_R^u\lsim 0.1$ and $\sin\theta_R^d\lsim 0.06$. The mixing angles for the left-handed fields are not independent and must satisfy $\tan\theta_L^u=\frac{m_t}{m_T}\tan\theta_R^u$ and
$\tan\theta_L^d=\frac{m_b}{m_B}\tan\theta_R^d$.

Concerning detection at the LHC, pair production processes dominated by QCD have the advantage of being model independent, with the new heavy quarks subsequently decaying into ordinary quarks and a gauge boson or a Higgs (see couplings in Eqs.~(\ref{chargedcurrents}), (\ref{neutralcurrents}) and (\ref{higgscurrent}) for our case).
A recent search \cite{atlaspair} yields observed lower limits on 
$T$ 
ranging between 715
and 950 GeV for all possible values of the branching ratios into the three decay modes
$T\rightarrow Wb$, $T\rightarrow Zt$ and $T\rightarrow Ht$.
Similarly,
for 
$B$
the
$B \bar B$ production implies that the limits range between 575 and 813 GeV for all possible values of the branching ratios into the three decay modes
$B\rightarrow Wt$, $B\rightarrow Zb$ and $B\rightarrow Hb$.
In these analyses, the above limits on mixing angles are applied since it is assumed that the new quarks mainly couple to the third generation. 
If they are allowed to mix with all standard model families, dedicated searches may be necessary.

The above mass bounds can be applied to our supersymmetric case if the light Higgs is a standard model-like Higgs particle and the decays of the fourth-family quarks involving non-standard model particles (such as e.g. the heavier Higgses or squarks) are negligible.
Otherwise, the new branching ratios should be taken into account implying a new phenomenology.
In addition, if the 
lepton-number-violating couplings $\lambda'$ of Eqs.~(\ref{superpotential1}) and (\ref{extratermss2}), which also violate $R_p$, are not small enough, they also could give rise to new channels modifying the single production of the new quarks, as well as their decay processes. The analysis of these possibilities is beyond the scope of this work, and we plan to cover it in a forthcoming publication~\cite{extended2}.
On the other hand, the new processes induced by the terms characteristics of the $\mn$, and violating also $R_p$, 
can be safely neglected because of the small value of $Y^\nu$.

\section{Conclusions}

In this work, in the framework of supersymmetry with right-handed neutrinos we have been able to reinterpret in a natural way the Higgs superfields as a fourth family of lepton superfields. 
From the theoretical viewpoint, this seems to be more satisfactory than the situation in usual supersymmetric models, where the Higgses are `disconnected' from the rest of the matter and do not have a three-fold replication.
Inspired by this interpretation of the Higgs superfields, we have also 
proposed the possible existence of a vector-like quark doublet representation in the low-energy supersymmetric spectrum.
These new quark superfields have the implication of a potentially rich 
phenomenology at the LHC. 

\section*{Acknowledgements}\label{ack}
We gratefully acknowledge J.A. Aguilar-Saavedra, A. Casas, J. Moreno and A. Uranga for useful discussions. 
The work of D.E. L\'opez-Fogliani  was supported by the Argentinian CONICET.
He acknowledges the hospitality of the IFT during whose stay this work was started.
The work of C. Mu\~noz was supported in part 
by the Programme SEV-2012-0249 `Centro de Excelencia Severo Ochoa'.
We also acknowledge the support of the Spanish grant FPA2015-65929-P (MINECO/FEDER, UE), and MINECO's Consolider-Ingenio 2010 Programme under grant MultiDark CSD2009-00064.

\end{document}